\documentclass{aa}

\usepackage{graphicx}
\usepackage{txfonts}
\usepackage{bm}
\usepackage[colorlinks=true,linkcolor=blue,citecolor=blue,urlcolor=blue]{hyperref}

\newcommand{\code}[1]{\textsc{#1}}
\newcommand{\deriv}[2]{\frac{\text{d}#1}{\text{d}#2}}
\newcommand{\tbts}{$T_b$--$T_s$\xspace}
\newcommand{\gcc}{\ensuremath{\rm\,g\,cm^{-3}}}

\begin{document}

\title{Neutron star envelopes with machine learning}

\subtitle{A single-hidden-layer neural network application}

\author{
    K. Kovlakas\inst{\ref{inst:k1},\ref{inst:k2}}\fnmsep\thanks{\email{kovlakas@ice.csic.es}} \and 
    D. De Grandis\inst{\ref{inst:k1},\ref{inst:k2}} \and
    N. Rea\inst{\ref{inst:k1},\ref{inst:k2}}
}

\institute{
    Institute of Space Sciences (ICE, CSIC), Campus UAB, Carrer de Magrans, 08193 Barcelona, Spain \label{inst:k1}
    \and 
    Institut d'Estudis Espacials de Catalunya (IEEC), Edifici RDIT, Campus UPC, 08860 Castelldefels (Barcelona), Spain \label{inst:k2}
}

\date{Received September 30, 20XX}

\abstract
{
Thermal and magneto-thermal simulations are an important tool for advancing understanding of neutron stars, as they allow us to compare models of their internal structure and physical processes against observations constraining macroscopic properties such as the surface temperature. A major challenge in the simulations is in modelling of the outermost layers, known as the envelope, exhibiting a drop of many orders of magnitude in temperature and density in a geometrically thin shell. This is often addressed by constructing a separate envelope model in plane-parallel approximation that produces a relation between the temperature at the bottom of the envelope, $T_b$, and the surface temperature, $T_s$. Our aim is to construct a general framework for approximating the \tbts relation that is able to include the dependencies from the strength and orientation of the magnetic field. We used standard prescriptions to calculate a large number of magnetised envelope models to be used as a training sample and employed single-hidden-layer feedforward neural networks as approximators, providing the flexibility, high accuracy, and fast evaluation necessary in neutron star simulations. We explored the optimal network architecture and hyperparameter choices and used a special holdout set designed to avoid overfitting to the structure of the input data. We find that relatively simple neural networks are sufficient for the approximation of the \tbts relation with an accuracy $\sim 3\%$.
The presented workflow can be used in a wide range of problems where simulations are used to construct approximating formulae.
}


\maketitle

\section{Introduction}
\label{sec:intro}

Neutron stars (NSs) are ultra-dense remnants of massive stellar cores and are observed as pulsars, thermally emitting compact objects, accreting X-ray binaries, gravitational-wave sources, gamma-ray bursts, and more. As the end points of the evolution of isolated stars with an initial mass $8\lesssim M_\odot\lesssim 30$ (depending on the metallicity) and with more diverse origins in interacting multiples (e.g. accretion induced collapse), they offer unique insights into stellar evolution theory \citep[e.g.][]{Taurs23}. Their extreme density, comparable or even exceeding the nuclear one, and strong magnetic fields ranging from several million to $\approx10^{15}\,$G probe physics in regimes unattainable in ground-based laboratories \citep[e.g.][]{Shapiro83}.

Many NSs have been detected to emit thermal emission in the soft X-ray band, corresponding to temperatures of $\approx 10^6\,$K \citep[e.g.][]{2024mbhe.confE..55R}; this can be compared with NS thermal evolution models as a means of accessing the internal structure and reactions occurring in the interior of the star \citep{1966CaJPh..44.1863T, 2004ApJS..155..623P, 2015SSRv..191..239P}.
When modelling the temperature throughout the star, very different scales must be considered. In particular, in the outermost layers, known as the `envelope', temperature and density drop by several orders of magnitude within mere meters \citep[e.g.][]{Haensel07}, making it problematic to treat them alongside the bulk of the star of $\approx10\rm\,km$ radius. To this end, envelope models are commonly studied in plane parallel and quasi-stationary approximation \citep[see the comprehensive review by][]{2021PhR...919....1B}.

Most notably, an envelope model is able to provide a \tbts relation, that is, an expression linking the temperature at the bottom of the envelope, $T_b$, to one at the surface, $T_s$. This in turn is key to link the observed temperature ($T_s$) to that controlling the physical processes in the interior ($T_b$). In particular, when solving numerically the NS thermal evolution equations in realistic, multidimensional setups, these envelope models are employed in order to impose a boundary condition on the surface temperature gradient \citep[e.g.][]{2019LRCA....5....3P}. However, this framework presents a practical nuisance: While the 1D parallel model is built using a given a value of $T_s$ as its initial condition (see Section~\ref{sec:physics}), the thermal evolution code updates the value of $T_b$. Hence, having an expression for the \tbts relation is necessary in all cases, as solving a 1D model every time the boundary condition is imposed is not only impractical, but impossible.
To this end, a variety of models and relations have been proposed in the literature. The case of a weakly magnetised, heavy element envelope was studied by \citet{1983ApJ...272..286G}, who provided a \tbts relation that is often taken as a reference. For more specific applications, a variety of more refined models taking into account the effect of the magnetic field, neutrino emission, and different chemical compositions are available in the literature \citep[e.g.][]{2003ApJ...594..404P,2015SSRv..191..239P,2016MNRAS.459.1569B}, typically in the form of ad hoc analytical formulae fitted to the models. However, as an increasing amount of details are included in the models, these expressions can become rather large and complicated to handle, especially if one needs to adapt them to include  further effects.

The above information highlights the necessity of a computational framework for generalizing the \tbts relation to consider additional parameters and construct an approximation that is accurate and fast. This is specifically important given the advent of 3D numerical codes \citep{2020ApJ...903...40D, 2021NatAs...5..145I, 2023MNRAS.518.1222D, 2024MNRAS.533..201A} since the \tbts relation is expected to be used at least once per time step for each point on the NS surface.

A promising avenue towards such a framework is the use of artificial neural networks (ANNs) as approximators of the \tbts relation. Firstly, ANNs can accommodate the inclusion of additional inputs without necessitating significant changes to the overall framework. More importantly, even though they can be slow in the training phase, their predictions are fast, involving simple algebraic operations that are highly optimised in modern computational software.
The flexibility and efficiency of ANNs have contributed to the widespread adoption of neural networks in astrophysics \citep[see review in][]{Smith23}. Initial steps to solve classification \citep[e.g.][]{Odewahn92} and regression problems \citep[e.g.][]{Collister04} were taken as early as the 1990s and 2000s. More complex neural network architectures, such as convolutional neural networks, have been used to obtain morphological classifications of galaxies from images (e.g. \citealt{Huertas15, Dominguez18}) and to provide alternatives to surface brightness profile fitting thousands of times faster than traditional methods (e.g. \citealt{Tuccillo18}.) These developments demonstrate how machine learning has evolved from performing relatively simple tasks (e.g. star–galaxy classification) to addressing increasingly complex problems, including constraining physics \citep[e.g.][]{Marino24} and employing explainable models to probe the behaviour of astrophysical sources \citep[e.g.][]{Oreste25}.

Multilayer feedforward networks are known to be universal approximators \citep[e.g.][]{Hornik89,Cybenko1989ApproximationBS} when given sufficient depth (number of hidden layers) and hidden units (neurons in hidden layers). The exact architecture of the network (number of hidden layers and units) and training configuration (e.g. initialisation, optimisation algorithm, number of epochs) depends on the investigated function and the structure of the training data. Interestingly, even the use of single-hidden-layer feedforward neural networks (SLFNs) can be sufficient for both univariate and multivariate functions \citep[e.g.][]{Cybenko1989ApproximationBS,Guliyev18}, especially when using a sigmoid activation function. This is particularly appealing for the application at hand, as it not only promises fast tuning and training but also faster predictions.

In this paper, we investigate the ability of SLFNs to approximate $T_s$ as a function of the magnetic field strength and orientation, bottom density and temperature, and chemical composition of the envelope (Section~\ref{txt:methods}). We study the optimal ANN architecture for future expansions of the models and provide trained networks for use by the community\footnote{\url{https://github.com/kkovlakas/nsenvelopes}} after having evaluated their approximation accuracy (Section~\ref{txt:results}). In Section \ref{txt:discussion}, we summarise our findings and show examples of \tbts relations, and we discuss the applicability of our approach to other fields. A direct application of the models and methods developed in this work can be found in \citet{DeGrandis25}, where a set of simulations of short-term magneto-thermal evolution in magnetars is presented.

\section{Methodology}
\label{txt:methods}

In this section we describe the physics input and the resulting simulation data. We also present the machine learning workflow towards a model approximating the \tbts relation.

\subsection{Physics input}\label{sec:physics}

We closely followed the formalism by \citet[see also \citealp{1977ApJ...212..825T}]{2007Ap&SS.308..353P}, which we present here for clarity of exposition. Namely, we describe the structure of the NS envelope, assuming a quasi-stationary state, as the set of equations for the gravitational potential, $\Phi$; local heat flux, $F_r$; temperature, $T$; and gravitational mass, $m$, enclosed in a sphere of radius, $r$:
\begin{equation}\begin{aligned}
    \deriv{\Phi}{\varpi}&=-\frac{1}{K_h}\frac{P}{\rho c^2},\\
    \frac{1}{r^2}\deriv{(r^2F_r)}{\varpi}&=\frac{P}{\rho g}\frac{Q_\nu}{K_hK_g}-2F_r\deriv{\Phi}{\varpi},\\
    \deriv{\ln T}{\varpi}&=\frac{3}{16} \frac{F_r}{\sigma T^4}\frac{\bar{\kappa} P}{g}\frac{1}{K_hK_g}-\deriv{\Phi}{\varpi},\\
    \deriv{r}{\varpi}&=-\frac{P}{\rho g}\frac{K_r}{K_hK_g},\\
    \deriv{m}{\varpi}&=-\frac{4\pi r^2 P}{g} \frac{K_r}{K_hK_g}.
\end{aligned}\label{eq:master}\end{equation}
Here, $\varpi=\ln P$ is the (natural) logarithm of the pressure, $P$; $\rho$ is the mass density; $Q_\nu$ is the neutrino emissivity; $\sigma$ is the Stefan-Boltzmann constant; $\bar \kappa$ is the opacity; and
\begin{equation}
    \begin{aligned}
    K_r &= (1-2Gm/rc^2)^{1/2}    \\
    K_h &= 1+ P/\rho c^2        \\
    K_g &= 1 + 4 \pi r^2 P/mc^2 \\
    g   &= Gm/(r^2K_r).              \\
    \end{aligned}
\end{equation}
These equations must be supplemented with an equation of state connecting the density and pressure $P=P(\rho,T,B,Z,A)$, which in general will also depend on the magnetic field, temperature, and composition ($Z$ and $A$ being the atomic and mass numbers of the plasma).
Within the notation used here, the opacity, $\bar\kappa$, is related to the thermal conductivity of the plasma, $\kappa$, as $\bar\kappa=16\sigma T^3/3\kappa\rho$, while the conductivity is affected by the magnetic field as
\begin{equation}
\kappa=\kappa_\parallel\cos^2\Theta+\kappa_\perp\sin^2\Theta,
\end{equation}
where $\Theta$ is the angle between the (local) magnetic field and the normal to the surface and $\kappa_{\parallel, \perp}$ are the components of the conductivity tensor related to the heat transport along or across field lines (themselves a function of $\rho$, $B$, $T$, and composition), respectively. The values of this quantity as well as the equation of state $P(\rho,\vec{B})$ were obtained through the latest version of the equation of state by \citet{2015SSRv..191..239P}, which is available online.\footnote{\url{https://www.ioffe.ru/astro/conduct/conduct.html}} We assumed a fixed chemical composition for the envelope, either of pure hydrogen (H) or pure iron (Fe).

This system is completed by the boundary conditions at the surface:
\begin{equation}
\begin{aligned}
    \Phi(\varpi_s) &= \ln K_r(\varpi_s), \quad
    &F_r(\varpi_s) &= \sigma T_s^4,      \\
    r(\varpi_s)    &= R_\star,           \quad
    &m(\varpi_s)   &= M_\star,    
\end{aligned}
\label{eq:boundary} 
\end{equation}
where the surface pressure logarithm, $\varpi_s$, is determined via the condition $\bar\kappa(\varpi_s)P(\varpi_s)/g(\varpi_s)=2/3$ \citep{1983ApJ...272..286G}. Therefore, an envelope model is defined once the stellar mass, $M_\star$, and radius, $R_\star$; surface temperature, $T_s$; and magnetic field (strength and direction) are defined.

The effect of neutrino emission, represented by the term $Q_\nu$ (which we described following the prescriptions from \citealp{2001PhR...354....1Y}), is to limit the maximum $T_s$ an envelope can have \citep{2015SSRv..191..239P}. In practice, when solving Equations~\ref{eq:master}, a large neutrino emission translates into a steep rise in the temperature. The cases in which this gradient makes the temperature increase up to unrealistic values ($\gtrsim 10^{10}\,$K) are interpreted as having too large an initial temperature $T_s$ to have a possible physical realisation and are therefore discarded.

We solved the system defined by Equations~\ref{eq:master}-\ref{eq:boundary} as an initial-values problem, with a standard fourth-order Runge-Kutta algorithm employing a step advance in $P$ in a geometric progression with constant 1\% increases. The integration proceeds up to an assigned density, $\rho_b$, the choice of which depends upon several factors. Namely, the integration should proceed throughout the region in which the steepest thermal gradient in the NS is found, which happens around the sensitivity strip, i.e. the layer where the radiative and electron opacity are comparable and the thermal gradient is largest \citep{1983ApJ...272..286G}. Typically, this happens at around $10^6\,$\gcc. However, the integration can be extended beyond this in order to remove the low density layers from the domain of (magneto-)thermal codes as much as possible; a customary choice is $\rho_b=10^{10}\,$g\,cm$^{-3}$. This value controls the time resolution of the (magneto-)thermal code itself, which cannot be shorter than the characteristic heat diffusion time across the envelope without breaking the assumption of quasi-stationarity onto which the \tbts relation is built (see the discussion in \citealp{2021PhR...919....1B}).
At $\rho_b=10^{10}\,$\gcc, this is $\approx 1\,$yr, which is adequate in most cooling simulations. However, when treating phenomena unfolding on shorter timescales, a lower value of $\rho_b$ should be used. In this work, we  concentrate on such lower values of $\rho_b$ since the motivation is to build envelope models
applicable to short-term simulations that can support high temperatures (this is most important for tackling outbursts; see \citealt{DeGrandis25}).
In particular, the term $Q_\nu$ increases with density so that models that are to be discarded when studying thick envelopes might still be physical for shallow ones (at which point the region at higher density does not have to be described under the constraint of stationarity).

In the following, we calculate models with two compositions, H and Fe, representing the extreme ends of NS envelope compositions, with other compositions falling between these cases. 
For simplicity, we assume a standard NS with $M_\star=1.4\,M_\odot$, $R_\star=10\rm\,km$. Different values of mass and radius can be accounted for, as the \tbts relation scales as the surface gravity $g_s^{1/4}$ \citep{1983ApJ...272..286G}, though an extension to account explicitly for this could be easily implemented within our formalism.

\subsection{Description of the simulation data}

Even though we have solved the full set of equations describing the structure of the envelope, in the following we focus on the profile of the temperature as a function of density $T(\rho)$
for a given set of physical parameters of the NS ($B$, $\Theta$, and $T_{\rm s}$). 
Moreover, in order to build envelope models corresponding to different $\rho_b$ values, we considered $\rho_b$ itself as another parameter. Thus the function $\mathcal{T}$ to be approximated is
\begin{equation}
    T_{\rm s} = \mathcal{T}(B, \Theta, T_i, \rho_i),
\end{equation}
where $i$ labels the integration grid. Collecting all simulation outputs gives
\begin{equation}\label{eq:Tsparams}
    T_{\rm s} = \mathcal{T}(B_j, \Theta_{j}, T_{ij}, \rho_{ij}).
\end{equation}
Here, $j$ enumerates the simulations, each corresponding to a temperature profile with multiple points (index $i$).

\begin{figure*}
    \centering
    \includegraphics[width=\textwidth]{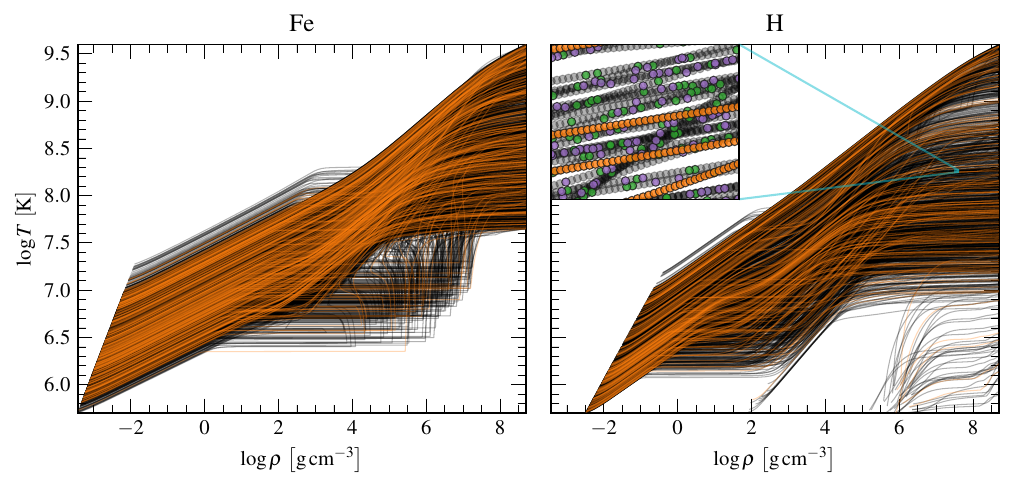}
    \caption{Profile curves from the NS envelope simulations for the Fe (\textit{left} panel) and H envelope (\textit{right} panel). Orange lines indicate the holdout curves used as an alternative test set. In the inset we show a zoom-in illustrating the random assignment of data points to the training (black), validation (green), and test (purple) sets from 90\% of the profiles as well as the holdout points (orange) comprised of whole curves (the remaining 10\%; see \S\ref{section:trainingcurves} for details.)}
    \label{fig:curves}
\end{figure*}

\begin{figure}
    \centering
    \includegraphics[width=\columnwidth]{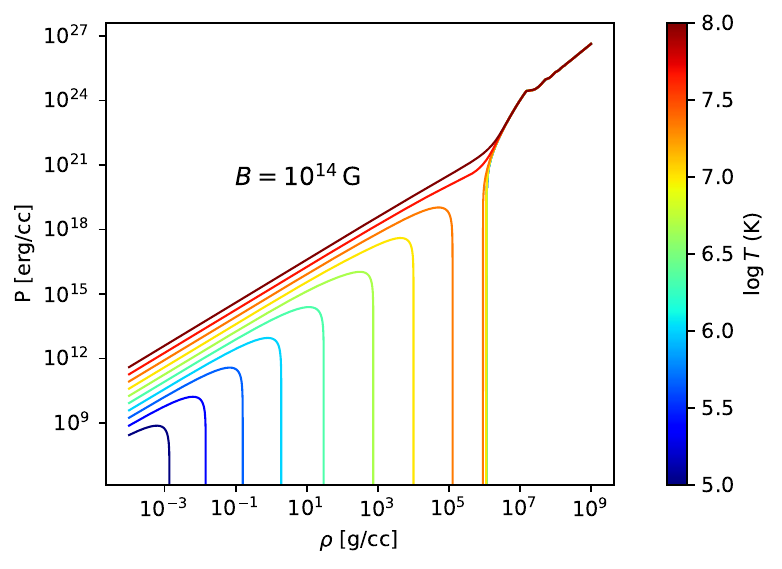}
    \caption{Equation of state by \citet{2013A&A...550A..43P} for a strong field and different temperature values in the case of iron composition. The gaps correspond to thermodynamically unstable regions (we note that the high density branches for all temperatures are almost superimposed and indistinguishable from one another).}
    \label{fig:eos}
\end{figure}

We ran a set of simulations randomly sampling the parameter space, namely selecting $B$ in the $[10^{9}, 10^{15}]\,$G range (with a log-uniform distribution), $\Theta$ uniformly in $\left[0, \pi/2\right]$, and $T_s$ log-uniformly in $[5\times10^5, 5\times10^7]\,\rm K$. Overall, our sample is constituted by 4221 profiles for the two compositions, which are shown in \autoref{fig:curves}. Interestingly, a fraction of the curves shows a behaviour quite unlike the others, as they exhibit plateaus (jump in density at the same $T$) and/or high density surfaces. These behaviours occur when the envelope presents a thermodynamically unstable or solid phase, respectively. These phases are due to the presence of the strong magnetic field, and for the parameter range studied here, they are the most important around $B\approx10^{14}\,$G. As an example, \autoref{fig:eos} shows the equation of state we adopted \citep{2013A&A...550A..43P} for a fixed field and different temperatures, which includes the transition to a thermodynamically unstable phase ($P<0$) at $\rho\lesssim10^6\,$\gcc. The `gap' in the equation of state is reflected in the density jumps in the envelope temperature profiles.

\subsection{Neural network architecture}

\begin{figure}
    \centering
    \includegraphics[width=\columnwidth]{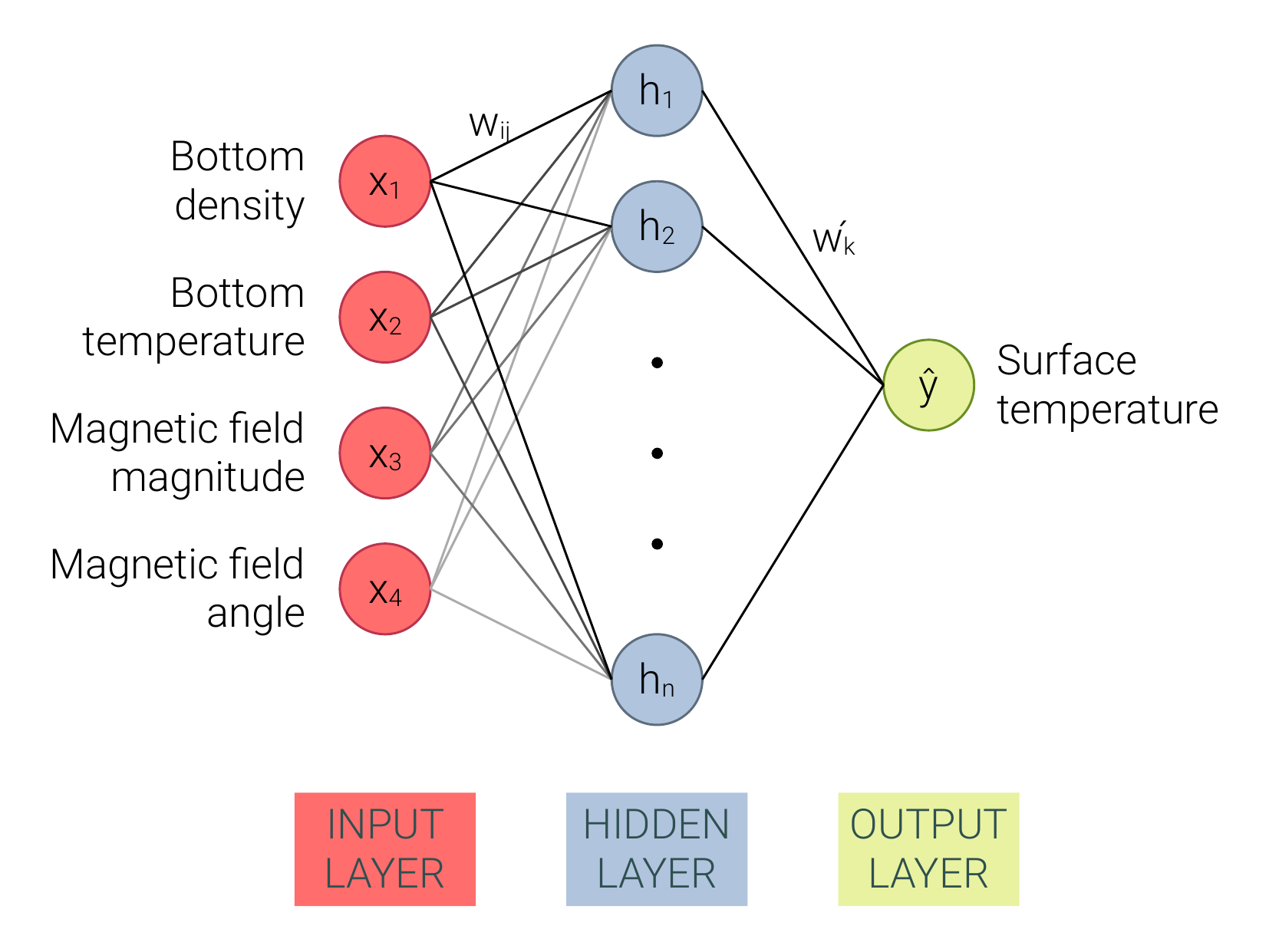}
    \caption{Single-hidden-layer feedforward network architecture used for approximating the surface temperature, $T_{\rm s}$. For $n$ hidden units ($h_j$), the total of variables is $6n+1$ (weights in each connection and biases in each unit including the output).}
    \label{fig:SLFN}
\end{figure}

The input layer consists of four units ($x_i$) corresponding to the four physical parameters in \autoref{eq:Tsparams}. They are the decimal logarithm of the density at the bottom of the envelope, $\log\rho$ (in units of $[\rm g\,cm^{-3}]$); the decimal logarithm of the corresponding temperature, $\log T$ (in units of $[\rm K]$); and the parameters of the magnetic field $\log B$ (in units of $[\rm G]$) and $\Theta$ (in radians).

The architecture is that of a shallow network having one hidden layer, as shown in \autoref{fig:SLFN}.
The hidden layer has $n$ units (the choice of this value is discussed in \S\ref{txt:besthype}) and is fully connected to the input layer, resulting in $4n$ weights $w_{ij}$ and $n$ biases $b_j$. The sigmoid activation function $S(x)=\left(1+e^{-x}\right)^{-1}$ is applied at each hidden unit, $h_j$,
\begin{equation}
h_j=S\left(b_j + \sum_{i=1}^{4} w_{ij} x_i\right).
\end{equation}
The output layer has one unit, corresponding to the target quantity, $\log T_{\rm s}$, and as it is fully connected to the hidden layer, it has $n$ weights ($w'_k$) and one bias ($b'$). As in standard regression neural networks, the linear activation function is used at the output ($f(x)=x$):
\begin{equation}
    \hat{y} = b' + \sum_{k=1}^{n} w'_k h_k. \label{eq:feedforward}
\end{equation}
Overall, the network contains a small number of trainable variables, $6n+1$.

We used standard prescriptions for the initialisation of the weights \citep{He15} and the optimisation method, that is, the \code{AMSGrad} variant \citep{Reddi19} of the \code{Adam} optimiser \citep{Kingma14}. For the loss function, we used the mean squared error (MSE) combined with L1 and L2 regularisation terms:
\begin{align}
    L &= \textrm{MSE} + R_{\rm L1} + R_{\rm L2} = \nonumber\\
      &= \frac{1}{N} \sum_{i=1}^{N} \left(y_i - \hat{y}_i\right)^2 
         + f \sum_m |w_m| + f\sum_m w_m^2.
\end{align}
Here, $y_i$ is the surface temperature of the profile containing the $i$-th data point, $\hat{y}_i$ is the network's prediction, while $w_m$ represent all the weights in the neural network. The regularisation terms, controlled by the hyperparameter $f$, are aimed at diminishing the effect of the weights, thus reducing the risk of overfitting.

\subsection{Data subsets and holdout curves}
\label{section:trainingcurves}

For each envelope model, we used four data sets. First, is the training set, which is the only one used to update the weights of the ANN. Second is the validation set used for the hyperparameter tuning (e.g. number of hidden units; see \S\ref{txt:besthype}) and for monitoring the validation error during training. The final sets are the test set and holdout curves for the final evaluation of the optimised and fully trained networks. In contrast to the test set, which comprises random points from different profiles, the holdout curves are the collection of all the integration points from a fraction of the simulated profiles. We note that the terms `test' and `holdout' are sometimes used interchangeably in the literature or to refer to different stages (final evaluation or validation). Here, they act as independent sets, which are unseen during training, tuning, or validation, and they are characterised by their different distributions in the feature space (random single points versus complete simulation profiles).

Specifically, we randomly selected 10\% of the profiles (see orange lines in \autoref{fig:curves}) as holdout curves. We merged the points of the remaining 90\% of the curves and randomly split them into the training (80\%), validation (10\%), and test (10\%) sets (72\%, 9\%, and 9\% of the total data, respectively). The inset in \autoref{fig:curves} demonstrates the distribution of the different sets in the $\log \rho-\log T$ feature space, where each sequence of points corresponds to the profile for a different combination of $B$ and $\Theta$. The training (black), validation (green), and test (purple) points are randomly distributed among the majority of the profiles, but the holdout points (orange) form full profiles.

\begin{figure*}
    \centering
    \includegraphics[width=\linewidth]{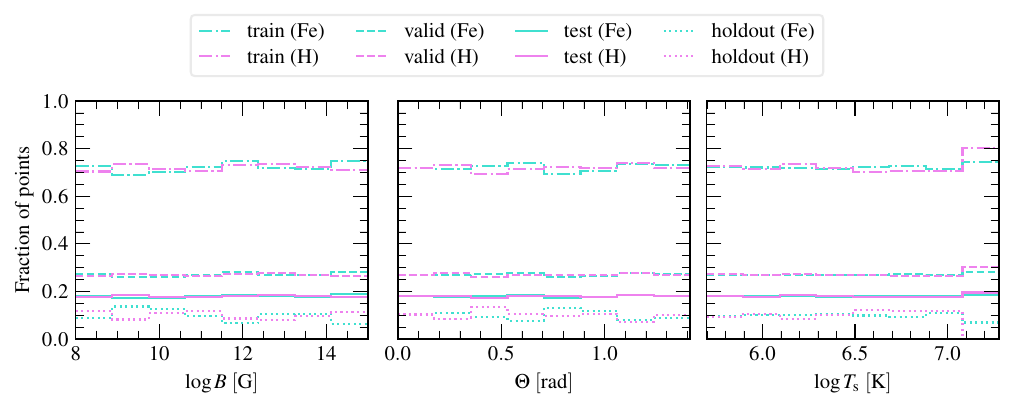}
    \caption{
    Fraction of data in the training (`train'), validation (`valid'), test, and holdout sets as a function of the input simulation parameters: magnetic field strength (\textit{left panel}) and angle (\textit{middle panel}), and the target variable surface temperature (\textit{right panel}) for the Fe (cyan) and H (violet) compositions. For better visibility, the fraction of the validation and test sets has been multiplied by three and two, respectively.
    }
    \label{fig:simparams}
\end{figure*}

Effectively, the neural networks will learn the profiles of the curves that are represented in the training set. Since the validation and test sets are made up of points from the same curves, there is a risk of overfitting the network to the specific values of $B$ and $\Theta$. The holdout curves serve as a way to test the ability of the final trained network to generalise, approximating $\mathcal{T}$ for unseen combinations of $(B, \Theta)$. In \autoref{fig:simparams} we show the fraction of data as a function of $B$, $\Theta$, and $T_{\rm s}$, demonstrating that the subsets are sampled fairly.

\subsection{Hyperparameter tuning}
\label{txt:besthype}

\begin{table}
\centering
\caption{Hyperparameters considered during the tuning phase.\label{tab:hype}}
\begin{tabular}{ll}
    \hline\hline
    Hyperparameters & Values \\\hline
    $n$, number of hidden layer units  & 256, 512, 1024, $\bm{2048}$, 4096
    \\
    $r$, learning rate                 & 0.001, $\bm{0.01}$, 0.1 \\
    $f$, regularisation factor         & $10^{-4}$, $10^{-5}$, $\bm{10^{-6}}$ \\\hline
\end{tabular}
\tablefoot{For each hyperparameter (first column) we explore various values (second column). The optimal values are marked in bold.}
\end{table}

After experimentation with different choices of hyperparameters (e.g. optimisers, weight initialisers), we identified three of them as determining the prediction accuracy, as well as the ranges that will be systematically explored. These are the learning rate, $r$; the number of units (neurons) in the hidden layer of the SLFN, $n$; and the regularisation factor, $f$ (see \autoref{tab:hype}). We used the \code{keras\_tuner} v.1.4.6 to perform a grid search on all 45 possible combinations, training the networks up to 100 epochs. Finally, we used 
the mean absolute error calculated on the validation set ($e_{\rm val}$) to measure the performance of the models, on which we later based our final choice.

\subsection{Training setup and early stopping}

We used \code{tensorflow} v.\ 2.15.0 for \code{Python} to compile the two SLFNs corresponding to the Fe and H envelope. We adopted the optimal hyperparameters (see \S\ref{txt:besthype}) and the same settings (e.g. optimiser, initialisation) as in the tuning phase.

Despite the small number of trainable variables ($10^{3-4}$ depending on the optimal number of units) with respect to the data ($\sim 7{\times}10^{6}$) and the use of regularisation, there is always a risk of overfitting. Moreover, it is quite possible that a small number of epochs is required for the task at hand, rendering prolonged training unnecessary. To address both issues, we used early stopping with patience to 30 epochs while monitoring the validation loss function. This means that during the training phase, the loss was calculated using the validation data that was not included in the learning process. If there was no improvement for 30 consecutive epochs, hinting at reaching the best performance, or overfitting, then the training stopped. Since during these 30 epochs the model may have been slightly overfitted to the training data, we restored the weights of the best performing snapshot of the network (30 epochs before the end).

Because of the use of early stopping, the final model depends on the random choice of the validation set. More importantly, the choice of restoring the weights results in a model with overrated performance if it is evaluated on the validation set. Therefore, we used the test set and holdout curves, which played no role in the construction of the model (learning or early stopping), to obtain an unbiased assessment of the performance on unseen data.

\section{Results}
\label{txt:results}

\subsection{Tuning}

In our investigation, since we are considering the two extreme cases of chemical composition, we expected them to be able to inform us about the best neural network architectures and scales for the estimation of $T_{\rm s}$ in further models with different compositions or physical assumptions.
For this reason, we combined the findings of the two grid searches. Furthermore, this increases the performance statistical size and reduces the risk of selecting hyperparameters due to statistical fluctuations in the performance metrics (e.g. mean absolute error).

Notably, for the H-envelope, the five best performing models out of the 45 trials had similar mean absolute errors ($0.020-0.022$) and corresponded to $r=0.01$, $f=10^{-6}$, and they all had different numbers of units (512, 4096, 256, 2048, 1024 in order of decreasing performance). We obtained similar results with the Fe-envelope models. The best performing models (mean absolute error between 0.015 and 0.19) had $f=10^{-6}$, with the majority having $r=0.01$, and the models had a varying number of units.

\begin{figure*}
    \centering
    \includegraphics[width=\linewidth]{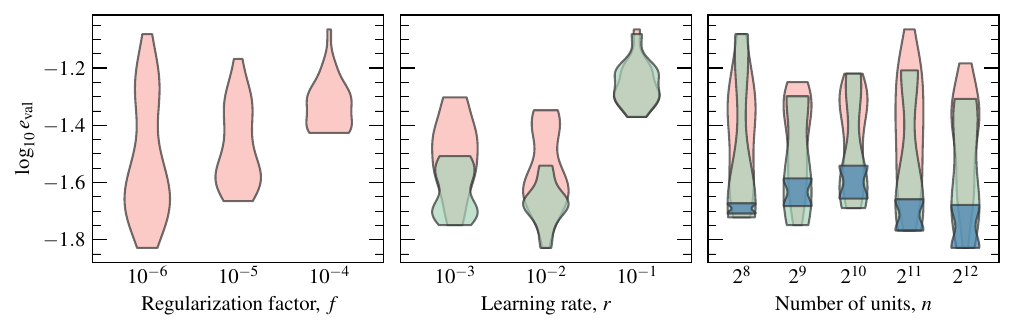}
    \caption{Distribution of the validation mean absolute errors of the models trained during the hyperparameter tuning. \textit{Left panel:} Dependence on the regularisation factor (pink), with $f=10^{-6}$ indicated as the optimal value. \textit{Middle panel}: Dependence on the learning rate (pink). The distribution highlights the trials with the optimal $f$ value (green) and indicates optimal performance for $r=0.01$. \textit{Right panel}: Dependence on the number of units (pink). The distribution highlights the trials with the optimal $f$ (green) and $r$ values (blue). The green violin plots focus on the models with the best performing regularisation factor, $f=10^6$, while the blue ones also focus on those with the best performing learning rate, $r=0.01$.}
    \label{fig:hyper}
\end{figure*}

In \autoref{fig:hyper} we show the distributions (with pink violin plots) of the validation mean absolute error of the trials as a function of the $f$, $r$, and $n$ (from left to right panel). As one can see in the $f$ panel, the best results are obtained for $f=10^{-6}$. For this reason, in the panels of $r$ and $n$, we also plot the subset of the trials with $f=10^{-6}$ (green violin plots). We observed that out of these models, the best performing ones had $r=0.01$. Consequently, in the last panel, we show the distribution of these models ($f=10^{-6}$ and $r=0.01$; blue violins), where we find that the best performance is obtained for 256, 2048, and 4096 units. This hints at the possibility that regularisation simplified the networks and resulted in a similar `effective' number of units.

Since the performance is similar when using different numbers of hidden units (possibly because of statistical fluctuations), we considered additional criteria for our final choice on this hyperparameter. A small number of units (e.g. 256) gives lower flexibility to the networks in case of more complex data for different compositions. On the other hand, a large $n$ would increase the computational complexity for predictions (feed-forward propagation). For this reason, we selected the intermediate value of $n=2048$. The hyperparameters we eventually adopted are highlighted in bold in \autoref{tab:hype}.

\subsection{Training and evaluation of performance}

\begin{figure}
    \centering
    \includegraphics[width=\columnwidth]{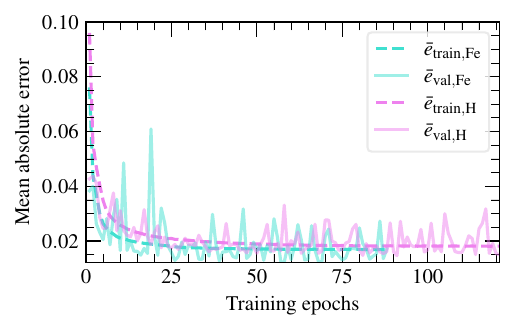}
    \caption{Improvement of the mean absolute error during the training of the SLFNs for the Fe (cyan) and H-envelope (violet) composition models, calculated on the training (dashed lines) and validation sets (solid lines). For illustrative purposes, the curves have been smoothed using a Savitzky-Golay filter with a window length equal to two.}
    \label{fig:training}
\end{figure}

In \autoref{fig:training} we show the training and validation mean absolute errors during the training of the two networks. The different number of epochs for the Fe (cyan) and H (violet) networks is a result of early stopping. We note that there is no sign of overfitting, and the models quickly converged after a few tens of epochs.

\begin{figure}
    \centering
    \includegraphics[width=\columnwidth]{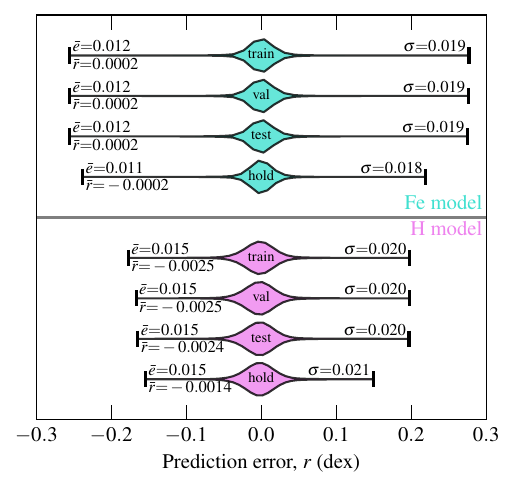}
    \caption{Distribution of the prediction errors, $r=y-\hat{y}$, of the final models for Fe (upper part) and H (lower part) in the different sets (from top to bottom): training (`train'), validation (`val'), test (`test'), and holdout curves (`hold'). The standard deviation of the prediction errors ($\sigma$), their mean value ($\bar{r}$; i.e. the mean bias of the predictions), and mean absolute values ($\bar{e}$; i.e. the mean absolute error) are shown in the tails of the violin plots.}
    \label{fig:residuals}
\end{figure}

Even with optimal average performance, it is important to ensure that the predictions are not systematically biased or skewed. To test this, we calculated the prediction errors (or residuals), $r_i=y_i{-}\hat{y}_i$, and their absolute values, $e_i=|r_i|$, for all four sets (training, validation, test, and holdout). The violin plots of the errors are shown in \autoref{fig:residuals}, marking the minimum and maximum values, their scatter ($\sigma$), mean value ($\bar{r}$), and the mean absolute error ($\bar{e}$).

The mean values of the prediction errors for both envelopes and all sets are $\lesssim 0.0025\rm\,dex$, which is significantly smaller than the scatter ($\lesssim 0.02\rm\,dex$), indicating that the predictions are not biased. The test and holdout mean absolute errors of the Fe model is $\sim 0.012\rm\,dex$, indicating a relative error smaller than $3\%$ in $T_{\rm s}$. For the H model, the mean absolute error is $\sim 0.015\rm\,dex$, resulting in a relative error of $\sim 3.5\%$ in $T_{\rm s}$.

\section{Discussion}
\label{txt:discussion}

\begin{figure*}
    \centering
    \includegraphics[width=\linewidth]{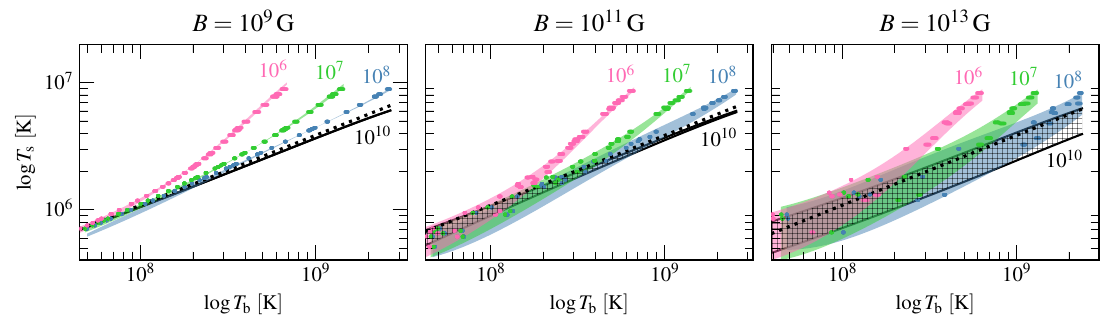}
    \caption{\tbts relations for different densities, $\rho_b$ (see annotated coloured text, in \gcc), and three magnetic field strengths (see titles) for the Fe envelope. The band corresponds to different values of the magnetic angle, $\Theta$ (lower values of $T_s$ for the same $T_b$ correspond to larger angles, i.e. more magnetic insulation).
    Using the same colours, we show all the points (circles) in the training dataset that correspond to similar magnetic field strengths and densities (difference smaller than $0.05\rm\,dex$), demonstrating the success of the model in learning the \tbts relation.
    The relations obtained by \citet{1983ApJ...272..286G} (dotted black line) and \citet{2015SSRv..191..239P} (hatched band) for $\rho_b=10^{10}\,\gcc$ are also plotted for comparison.}
    \label{fig:tbts}
\end{figure*}

\begin{figure*}
    \centering
    \includegraphics[width=\linewidth]{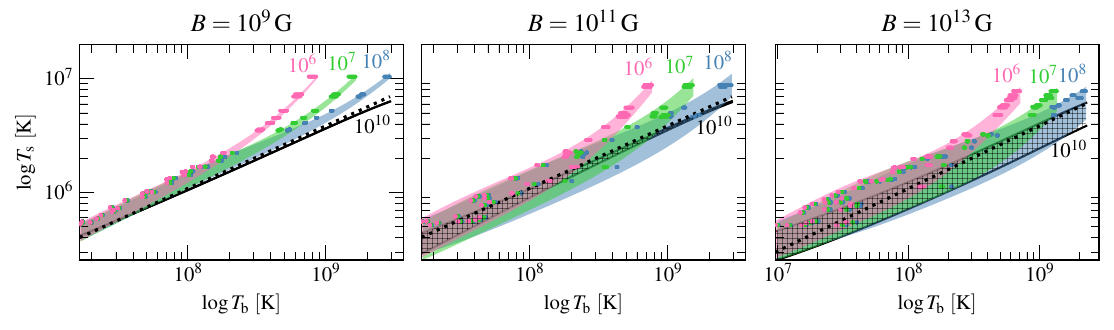}
    \caption{Same as \autoref{fig:tbts} but for the H envelope. The hatched band in this case is the model from \citet{2003ApJ...594..404P}.}
    \label{fig:tbts_H}
\end{figure*}

The present work showcases the step-by-step construction of an approximator of the \tbts relation that is directly applicable to numerical codes for NS evolution. The same recipe can be followed by studies adopting different models of NS envelopes without further extensive exploration of the hyperparameter space.

In Figures~\ref{fig:tbts} and \ref{fig:tbts_H}, we show the predictions of the Fe and H envelope, respectively, for different magnetic field strengths and inclinations and bottom densities. We also overplot (circles) the training data that correspond to the respective field values (with a tolerance of $0.05\rm\,dex$ absolute difference since the training data correspond to simulations with randomly sampled magnetic field strengths). We find that the model predictions follow the distribution of the points, and despite the small number of points, the bands are smooth, indicating that the network has learned the trends in the \tbts relation.
The overall trends are also compatible with the fits found in the literature and computed for $\rho_b=10^{10}\,$\gcc. We did not extend our curves up to this density value because those with a high $T_s$ would have been rendered unphysical by neutrino emission (which is not necessarily the case when those layers are treated outside of the plane-parallel approximation). Nevertheless, we checked that in the cases in which such $\rho_b$ could be reached our models follow the existing \tbts\ relation fits from the literature well within their error.

We find that a small neural network having 2048 neurons (units) in a single hidden layer is sufficient for reaching an accuracy of $<0.015\rm\,dex$ ($<3.5\%$) in the case of the Fe and H envelopes.
The use of the sigmoid activation function avoids artefacts and numerical instabilities in simulations and is compatible with numerical solvers requiring smooth functions.
The result is a fast and infinitely differentiable approximator that can be loaded in modern computational frameworks (e.g. in Python) or even manually built with ease to interact with pre-existing codes (e.g. in Fortran). In particular, we implemented our model in the Fortran code by \citet{2019LRCA....5....3P}, finding a limited impact on performance (total runtime increased by $1$--$5\%$) with respect to the use of analytical expressions from the literature \citep[specifically, the one proposed in][valid only for a single $\rho_b$]{2015SSRv..191..239P} on a standard laptop running on an Intel i7 processor.

The weak dependence of the mean absolute error on the number of units in the hidden layer indicates that the two models are relatively simple: Many of the weights are close to zero, and a small number of neurons are activated. Consequently, different models with wider parameter ranges or additional functional dependence on other NS properties may be accommodated with the same adopted hyperparameter values.

Shallow-learning approaches with neural networks are not new in astrophysics \citep[e.g.][]{Silva11}, but to our knowledge this is the first study applying them in data that represent `evolution tracks', as well as considering an additional test dataset (holdout curves) to detect potential overfitting due to the ordered structured of the training data. Similar networks and holdout curve sets can be used in other disciplines where fast approximations are desired involving evolution in space or time, such as stellar or planetary structure; single, binary, or triple population synthesis; metallicity-dependent star-formation histories of galaxies. The code of the presented framework is provided in a public \code{GitHub} repository,\footnote{\url{https://github.com/kkovlakas/nsenvelopes}}
, and we have included instructions as well as the ready-for-use trained models.

\begin{acknowledgements}
KK is supported an ICE Fellowship under the program Unidad de Excelencia Maria de Maeztu (CEX2020-001058-M).
KK, DDG and NR are supported by the Catalan grant SGR2021-01269 (PI: Graber/Rea), DDG by a Juan de la Cierva fellowship (JDC2023-052264-I), and DDG and NR are also supported by the European Research Council (ERC) via the Consolidator Grant `MAGNESIA' (No. 817661) and the Proof of Concept `DeepSpacePulse' (No. 101189496).
\end{acknowledgements}

\bibliographystyle{aa}
\bibliography{main}

\end{document}